\documentstyle[aps,prl,12pt,psfig]{revtex}
\begin{document}

\begin{center}

\today

{\large \bf Infrared reflectivity of pure and doped  CuGeO$_3$ \\
}
\bigskip

A. Damascelli$^a$, D. van der Marel$^a$, F. Parmigiani$^b$,
G. Dhalenne$^c$, A. Revcolevschi$^c$\\

{\it 
$^a$Solid State Physics Laboratory, University
 of Groningen, Nijenborgh 4, 9747 AG Groningen, The Netherlands\\
 $^b$INFM and Dipartimento di Fisica, Politecnico 
di Milano, Piazza Leonardo da Vinci, 32-20133 Milano, Italy\\
 $^c$Laboratoire de Chimie des Solides, Universit$\acute{e}$ de Paris-sud, 
B$\hat{a}$timent 414, F-91405 Orsay, France
}

\end{center}
\bigskip

\noindent

We investigated the far- and mid-infrared  reflectivity (20-6000 cm$^{-1}$) of several
  Cu$_{1-\delta}$Mg$_{\delta}$GeO$_3$  ($\delta$=0,$\,$0.01) and 
CuGe$_{1-x}$Si$_{x}$O$_3$ ({\em x}=0,$\,$0.007,$\,$0.05,$\,$0.1)
single crystals. The b-axis and c-axis optical response 
is presented for different temperatures between 4 K and 300 K. Moreover, a full group theoretical 
analysis of the lattice vibrational modes of CuGeO$_3$ in the high temperature undistorted 
phase as well as in the low temperature spin-Peierls phase is reported and compared with the experimental 
results. We observe the activation of zone boundary phonons along the b axis of the crystal 
below the spin-Peierls transition temperature.

\bigskip

\noindent
keywords:  CuGeO$_3$; spin-Peierls transition; phonons
\vspace{1 cm}

\noindent
Andrea Damascelli\\
Laboratory of Solid State Physics\\
University of Groningen\\
Nijenborgh 4, 9747 AG Groningen\\
The Netherlands\\
phone: +31-50-363 4922 \hspace{0.5 cm}fax:  363 4825\\
email: damascel@phys.rug.nl
\newpage
%\twocolumn

\section {Introduction}

Since CuGeO$_3$ has been recognized, on the basis of magnetic susceptibility measurements \cite{hase}, 
as the first inorganic compound showing a spin-Peierls transition (SP), it has attracted the attention 
of many scientists. In particular, one of the main points of interest was to observe the structural 
phase transition, occurring in conjunction with  the formation of  a non-magnetic singlet ground state, 
both processes being driven by the antiferromagnetic interaction in the one-dimensional Cu-O chains. 
In  fact, many papers on x-ray and neutron scattering experiments have been published during the last 
few years, reporting not only the dimerization of the Cu-Cu pairs along the c axis \cite{pouget} but 
also the displacement of the Ge and O atoms in the a-b plane \cite{hirota,braden}. Optical 
techniques are also very useful in investigating magnetic and/or structural phase transitions. Although a 
considerable amount of Raman scattering \cite{kuroe,jandl,goni} and  infrared spectroscopy \cite{popovic,massa,loosdrecht2} work has been devoted to the undistorted and the SP phase of CuGeO$_3$, an  investigation of the full temperature dependence of the infrared active phonons 
has not been reported so far.

In this paper, infrared reflectivity measurements on pure and doped CuGeO$_3$  single crystals, 
for temperatures ranging from 4 K to 300 K, are presented and compared with the results of a group 
theoretical analysis performed for the two different crystal structures of CuGeO$_3$ .

\section {Group theoretical analysis}

At room temperature CuGeO$_3$ has an orthorhombic crystal structure with lattice parameters
 a=4.81 \AA, b=8.47 \AA\    and c=2.941 \AA\      and space group Pbmm (x$\|$a, y$\|$b, z$\|$c) or,
 equivalently, Pmma (x$\|$b, y$\|$c, z$\|$a) in standard setting \cite{vollenkle}. The building 
 blocks of the structure are
 edge-sharing CuO$_6$ octahedra and corner-sharing GeO$_4$ tetrahedra stacked along the 
c axis of the crystal and resulting in Cu$^{2+}$ and Ge$^{4+}$ chains parallel to the c axis. These 
chains are linked together via the O atoms and form layers parallel to the b-c plane weakly 
coupled along the a axis (Fig.~1). The unit cell contains 2 formula units of CuGeO$_3$ (Fig.~2), 
with site group C$_{2\rm{h}}^{\rm{y}}$ for the 2 Cu atoms, C$_{2\rm{v}}^{\rm{z}}$ for the 2 Ge 
and the 2 O(1) atoms and C$_{\rm{s}}^{\rm{xz}}$ for the 4 O(2) atoms (where O(2) denotes the O 
atoms linking the chains together) \cite{hirota,braden}. A group theoretical analysis can be 
performed, working in standard orientation, to obtain the number and the symmetry of the lattice 
vibrational modes. Following the nuclear site group analysis method extended to crystals \cite{rousseau}, 
the contribution of each occupied site to the total irreducible representation of the crystal is:\\
\\
$\Gamma_{\rm{Cu}}$=A$_{\rm{u}}$+2B$_{\rm{1u}}$+B$_{\rm{2u}}$+2B$_{\rm{3u}}$~,
\\
$\Gamma_{\rm{Ge+O(1)}}$=2[A$_{\rm{g}}$+B$_{\rm{1u}}$+B$_{\rm{2g}}$+
B$_{\rm{2u}}$+B$_{\rm{3g}}$+B$_{\rm{3u}}$]~,
\\
$\Gamma_{\rm{O(2)}}$=2A$_{\rm{g}}$+A$_{\rm{u}}$+B$_{\rm{1g}}$+2B$_{\rm{1u}}$+
2B$_{\rm{2g}}$+B$_{\rm{2u}}$+B$_{\rm{3g}}$+2B$_{\rm{3u}}$~.\\
\\
Subtracting the silent modes (2A$_{\rm{u}}$) and the acoustic modes 
(B$_{\rm{1u}}$+B$_{\rm{2u}}$+B$_{\rm{3u}}$), the irreducible representation of the optical vibrations 
in standard setting (Pmma), is:\\
\\
$\Gamma$=4A$_{\rm{g}}$(aa,bb,cc)+B$_{\rm{1g}}$(bc)+4B$_{\rm{2g}}$(ab)
+3B$_{\rm{3g}}$(ac)+5B$_{\rm{1u}}$(E$\|$a)+3B$_{\rm{2u}}$(E$\|$c)+
5B$_{\rm{3u}}$(E$\|$b)~.\\
\\
This corresponds to an expectation of 12 Raman active modes (4A$_{\rm{g}}$+B$_{\rm{1g}}$
+4B$_{\rm{2g}}$+3B$_{\rm{3g}}$) and 13 infrared active modes (5B$_{\rm{1u}}$
+3B$_{\rm{2u}}$+5B$_{\rm{3u}}$) for CuGeO$_3$, in agreement with the calculation done by Popovi\'c et 
al. \cite{popovic}. 

At temperatures lower than T$_{\rm{SP}}$ the crystal structure is still orthorhombic, but with lattice 
parameters a'=2$\times\rm{a}$, b'=b and c'=2$\times\rm{c}$ and space group Bbcm (x$\|$a, y$\|$b, z$\|$c) 
or , equivalently, 
Cmca (x$\|$c, y$\|$a, z$\|$b)  in standard setting \cite{hirota,braden}. The distortion of the lattice 
taking place at the phase transition (Fig.~2) is characterized by the dimerization of  Cu-Cu pairs along the 
c axis (dimerization out of phase in neighboring chains), together with a rotation of the GeO$_4$ 
tetrahedra around the axis defined by the O(1) sites (rotation opposite in sense 
for neighboring tetrahedra). Moreover, the O(2) sites of the undistorted structure split in an equal 
number of O(2a) and O(2b) sites, distinguished by the distances O(2a)-O(2a) and O(2b)-O(2b) shorter and 
larger than O(2)-O(2) \cite{braden}, respectively. The SP transition is also characterized (Fig.~2) by 
a doubling of the unit cell (corresponding to a doubling of the degrees of freedom from 30 to 60); the 
site groups in the new unit cell are: C$_{2}^{\rm{x}}$ for Cu, C$_{2}^{\rm{y}}$ for O(1) and 
C$_{\rm{s}}^{\rm{yz}}$ for Ge, O(2a) and O(2b) \cite{braden}.  Repeating the group theoretical 
analysis we obtain for the contributions to the total irreducible representation:\\
\\
$\Gamma_{\rm{Cu}}$=A$_{\rm{g}}$+A$_{\rm{u}}$+2B$_{\rm{1g}}$+2B$_{\rm{1u}}$+
2B$_{\rm{2g}}$+2B$_{\rm{2u}}$+B$_{\rm{3g}}$+B$_{\rm{3u}}$~,
\\
$\Gamma_{\rm{Ge+O(2a)+O(2b)}}$=3[2A$_{\rm{g}}$+A$_{\rm{u}}$+B$_{\rm{1g}}$+
2B$_{\rm{1u}}$+B$_{\rm{2g}}$+2B$_{\rm{2u}}$+2B$_{\rm{3g}}$+B$_{\rm{3u}}$]~,
\\
$\Gamma_{\rm{O(1)}}$=A$_{\rm{g}}$+A$_{\rm{u}}$+2B$_{\rm{1g}}$+2B$_{\rm{1u}}$+
B$_{\rm{2g}}$+B$_{\rm{2u}}$+2B$_{\rm{3g}}$+2B$_{\rm{3u}}$~.\\
\\
The irreducible representation of the optical vibrations of CuGeO$_3$ in the SP phase in standard 
setting (Cmca), is:\\
\\
$\Gamma_{\rm{SP}}$=8A$_{\rm{g}}$(aa,bb,cc)+7B$_{\rm{1g}}$(ac)+
6B$_{\rm{2g}}$(bc)+9B$_{\rm{3g}}$(ab)+9B$_{\rm{1u}}$(E$\|$b)+
8B$_{\rm{2u}}$(E$\|$a)+5B$_{\rm{3u}}$(E$\|$c)~.\\
\\
Therefore 30 Raman active modes (8A$_{\rm{g}}$+7B$_{\rm{1g}}$
+6B$_{\rm{2g}}$+9B$_{\rm{3g}}$) and 22 infrared active modes (9B$_{\rm{1u}}$
+8B$_{\rm{2u}}$+5B$_{\rm{3u}}$) are expected  for CuGeO$_3$ in the SP phase, all the additional vibrations 
being zone boundary modes activated by the folding of the Brillouin zone.

To compare the results obtained for the undistorted and the SP phase of  CuGeO$_3$ it is better to rewrite 
the irreducible representations $\Gamma$ and $\Gamma_{\rm{SP}}$ into Pbmm and Bbcm settings, respectively, 
because both groups are characterized by: x$\|$a, y$\|$b and z$\|$c. This can be done by permuting 
the  (1g,2g,3g) and (1u,2u,3u) indices  in such a way that  it corresponds  to the permutations of the 
axis relating Pmma to Pbmm and Cmca to Bbcm. Therefore, the irreducible representations of the optical 
vibrations of CuGeO$_3$,  for T$>$T$_{\rm{SP}}$ (Pbmm) and T$<$T$_{\rm{SP}}$ (Bbcm), respectively, are:\\
\\
$\Gamma^{\prime}$=4A$_{\rm{g}}$(aa,bb,cc)+
4B$_{\rm{1g}}$(ab)
+3B$_{\rm{2g}}$(ac)+B$_{\rm{3g}}$(bc)+3B$_{\rm{1u}}$(E$\|$c)+
5B$_{\rm{2u}}$(E$\|$b)+5B$_{\rm{3u}}$(E$\|$a)~,\\
\\
\\
$\Gamma^{\prime}_{\rm{SP}}$=8A$_{\rm{g}}$(aa,bb,cc)+9B$_{\rm{1g}}$(ab)+7B$_{\rm{2g}}$(ac)
+6B$_{\rm{3g}}$(bc)+5B$_{\rm{1u}}$(E$\|$c)+9B$_{\rm{2u}}$(E$\|$b)+8B$_{\rm{3u}}$(E$\|$a)~.\\
\\
It is now evident that the number of infrared active phonons  is expected to increase from 5 to 8,  
5 to 9 and 3 to 5 for  light polarized along the a, b and c axes, respectively.

\section{Experimental}

We investigated the far- and mid-infrared  reflectivity (20-6000 cm$^{-1}$) of several
  Cu$_{1-\delta}$Mg$_{\delta}$GeO$_3$  ($\delta$=0,$\,$0.01) and 
CuGe$_{1-x}$Si$_{x}$O$_3$ ({\em x}=0,$\,$0.007,$\,$0.05,$\,$0.1)
single crystals. These high-quality single crystals were grown 
from the melt by a floating zone technique \cite{revcolevschi}. Samples with dimensions of 
1$\times$3$\times$6 mm$^3$ were aligned by conventional Laue diffraction and mounted in a liquid He 
flow cryostat to study the temperature dependence of the optical properties between 
4 K and 300 K. The reflectivity measurements were performed with a Fourier transform 
spectrometer (Bruker IFS 113v), operating in near normal incidence configuration with polarized
 light in order 
to probe the optical response of the crystals along the b and the c axes. The absolute reflectivities 
were obtained by calibrating the data acquired on the samples against a gold mirror.

\section{Results}

\subsection{Pure CuGeO$_3$}
The c- and b-axis reflectivity spectra of CuGeO$_3$  in the undistorted phase are shown in Fig.~3, for 
two different temperatures. The data are shown up to 1000 cm$^{-1}$ which covers  the full phonon spectrum. 
The spectra are typical of an insulating material. Three phonons are detected  along the c axis 
($\omega_{\rm{TO}}\approx$\,167, 528 and 715 cm$^{-1}$ for T=15 K), and five along the b axis 
($\omega_{\rm{TO}}\approx$\,48, 210, 286, 376 and 766  cm$^{-1}$ for T=15 K), in agreement with 
the calculation presented in section II. The structure in Fig.~3(a) between 200 and 400 cm$^{-1}$ 
is due to a leakage of the modes polarized along the b axis and the feature at approximately 630 
cm$^{-1}$  in Fig.~3(b) is a leakage of a mode polarized along the a axis \cite{popovic}. 

In Fig.~4 the reflectivity measured at 4 K in the SP phase is compared with the data obtained just 
above T$_{\rm{SP}}$=14 K. Whereas for E$\|$c the spectra are exactly identical, a new feature is 
detected in the SP phase  at 800 cm$^{-1}$  for E$\|$b, as clearly shown in the inset of  Fig.~4(b). 
This line (that falls in the frequency region of high reflectivity for the phonon at 766  cm$^{-1}$ 
and therefore shows up mainly for its absorption) can be, in our opinion, interpreted as a folded mode 
due to the SP transition. The reason for  not observing all phonons predicted from the group theoretical 
analysis is probably the small value of the atomic displacements involved in the SP transition, with a 
correspondingly small oscillator strength of zone boundary phonons.

\subsection{Doped CuGeO$_3$}
The reflectivity data acquired on the Si doped samples for E$\|$c and E$\|$b are shown in Fig.~5 and 6, 
respectively. Some new features, due only to the substitution of Ge with the lighter Si and not directly 
related to the SP transition, are observable: new phonon peaks at 900 cm$^{-1}$  along the c axis (Fig.~5) 
and at 500 and 960 cm$^{-1}$  along the b axis (Fig.~6). Moreover, the more complicated line shape and 
the reduction of the oscillator strength of the high frequency phonons indicate a strong Ge (Si) 
contribution to these modes, mainly due to O vibrations. At low temperatures we observe the mode at 
800 cm$^{-1}$  for E$\|$b only for the lowest Si concentration, as shown in the inset of Fig.~6(b), 
where the 4 K and 11 K data are compared. We conclude that up to 0.7\% Si doping the SP transition is 
still present with T$_{\rm{SP}}<$11 K, whereas for 5\% and 10\% Si concentrations no indication of the 
transition could be found in our spectra. 

The results obtained on the Mg-doped sample are plotted in Fig.~7  and 8 for E$\|$c and E$\|$b, 
respectively. Clearly Mg doping is affecting the optical response of  CuGeO$_3$  less than Si doping. 
A new phonon due to the mass difference between Cu and Mg is present in the c-axis spectra at 695 
cm$^{-1}$, as clearly shown in the inset of  Fig.~7(b),  for T=4 K. Moreover, we clearly observe 
for E$\|$b [see inset of  Fig.~8(b)],  the 800 cm$^{-1}$  feature related to the SP transition. 
On the one hand, for the 1\% Mg-doped sample, T$_{\rm{SP}}$ seems to be lower than in pure CuGeO$_3$; 
on the other hand the structural deformation is not as strongly reduced as in the 0.7\% Si-doped 
sample, as can be deduced from the comparison between  the insets of  Fig.~6(b)  
and  Fig.~8(b).

\section{Conclusions}
In summary, we have investigated the infrared reflectivity of pure and Mg- and Si-doped CuGeO$_3$ 
single crystals, for E polarized parallel to the b  and c axes of the crystals and temperatures ranging 
from 4 K to 300 K. The results of a group theoretical calculation for the undistorted and distorted 
structures of CuGeO$_3$  are presented and compared with the experimental data. A feature reflecting the 
SP transition has been observed in the phonon spectrum and interpreted as a folded zone boundary mode. 

\section {Acknowledgments}

We gratefully acknowledge M. Mostovoi and D.I. Khomskii 
for stimulating discussions. We thank K. Schulte, A.-M. Janner and M. Gr\"uninger 
 for many useful comments and  Z. Tomaszewski for technical help. 
This investigation was supported by the Netherlands Foundation for
Fundamental Research on Matter (FOM) with financial aid from
the Nederlandse Organisatie voor Wetenschappelijk Onderzoek (NWO).

%\begin{thebibliography}{99}

\newpage

{\bf Figure captions}

\begin{figure}[htb]
%\vspace{2.5cm}
%\centerline{\psfig{figure=ascfig1.eps,width=7cm,clip=}}
 \caption{Crystal structure of CuGeO$_3$ in the high temperature (T=300 K) undistorted phase.}
\end{figure}

\begin{figure}[htb]
%\centerline{\psfig{figure=ascfig2.eps,width=7cm,clip=}}
 \caption{Conventional unit cell of CuGeO$_3$ in the undistorted 
(top) and SP phase (bottom). For clarity purpose the ion displacements
due to the SP transition have been enlarged by a factor of 30.
}
\end{figure}

\begin{figure}[htb]
%\centerline{\psfig{figure=ascfig3.eps,width=7cm,clip=}}
 \caption{Reflectivity of a single crystal of pure CuGeO$_3$ as a function of wavenumber 
at two different temperatures (300 K and 15 K)   in the undistorted phase. The spectra 
are shown for light  polarized along the c axis (a) and the b axis (b) of the crystal.
}
\end{figure}

\begin{figure}[htb]
%\centerline{\psfig{figure=ascfig4.eps,width=7cm,clip=}}
 \caption{Comparison between reflectivity spectra measured in the SP phase at 4 K 
(solid line) and just before the SP transition at 15 K (circles) on a pure single-crystal of
CuGeO$_3$. For light polarized along the c axis (a) no difference is found across the phase
 transition whereas for  light polarized along the b axis (b) a new feature appears 
at 800 cm$^{-1}$ (as clearly shown in the inset).
}
\end{figure}

\begin{figure}[htb]
%\centerline{\psfig{figure=ascfig5.eps,width=7cm,clip=}}
 \caption{C-axis reflectivity of Si-doped single crystals of CuGeO$_3$  
for different silicon concentrations. The spectra are presented as a function of 
wavenumber for T=300 K (a) and T=4 K (b).
}
\end{figure}

\begin{figure}[htb]
%\centerline{\psfig{figure=ascfig6.eps,width=7cm,clip=}}
 \caption{B-axis reflectivity of Si-doped single crystals of CuGeO$_3$  
for different silicon concentrations. The spectra are presented as a function of 
wavenumber for T=300 K (a) and T=4 K (b). The 800 cm$^{-1}$ excitation 
activated by the SP transition is still observable for 0.7\% Si-doping by comparing (see inset) 
the 4 K (solid line) and the 11 K (circles) data.
}
\end{figure}

\begin{figure}[htb]
%\centerline{\psfig{figure=ascfig7.eps,width=7cm,clip=}}
 \caption{C-axis reflectivity of pure and 1\% Mg-doped single crystals of CuGeO$_3$. 
The spectra are presented as a function of wavenumber for T=300 K (a) and T=4 K (b). 
The inset shows an enlarged view of the frequency region around 700 cm$^{-1}$ for  the 
data obtained on the pure (circles) and Mg-doped (solid line) samples, at T=4 K. The additional 
peak observed for Mg doping is due to the mass difference between Cu and Mg and it is
not related to the SP transition.
}
\end{figure}

\begin{figure}[htb]
%\centerline{\psfig{figure=ascfig8.eps,width=7cm,clip=}}
 \caption{B-axis reflectivity of pure and 1\% Mg-doped single crystals of CuGeO$_3$. 
 The spectra are presented as a function of  wavenumber for T=300 K (a) 
and T=4 K (b). For the Mg-doped sample the 800 cm$^{-1}$ excitation activated by the SP transition 
is clearly observable  in the inset, where the 4 K (solid line) and 
the 14 K (circles) data are presented.
}
\end{figure}

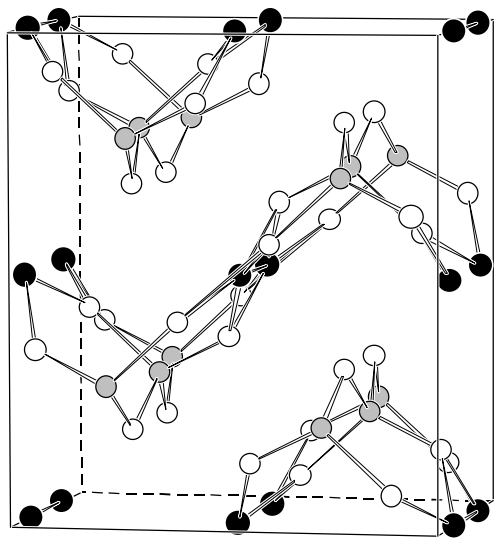
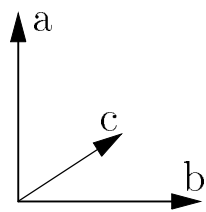
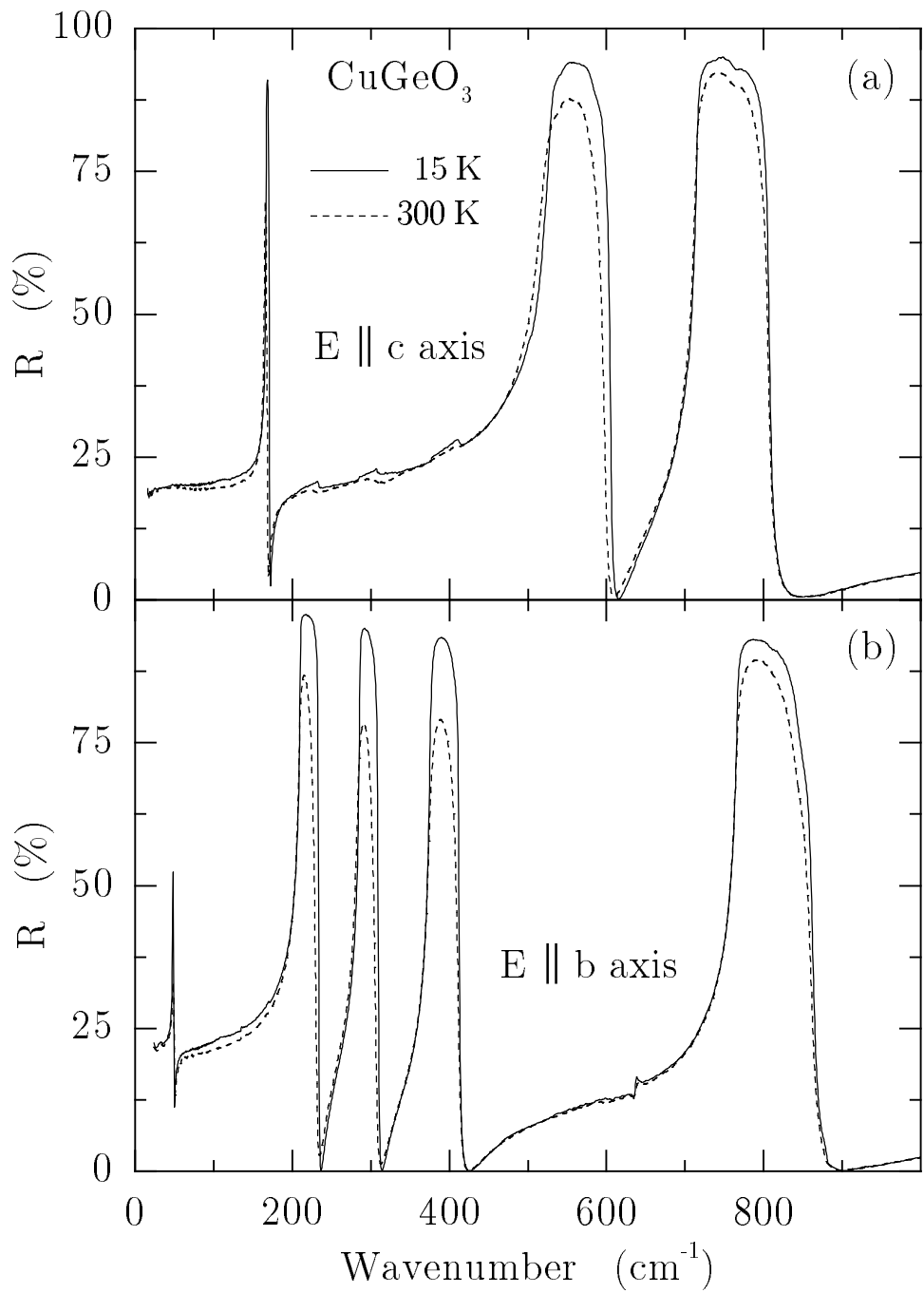
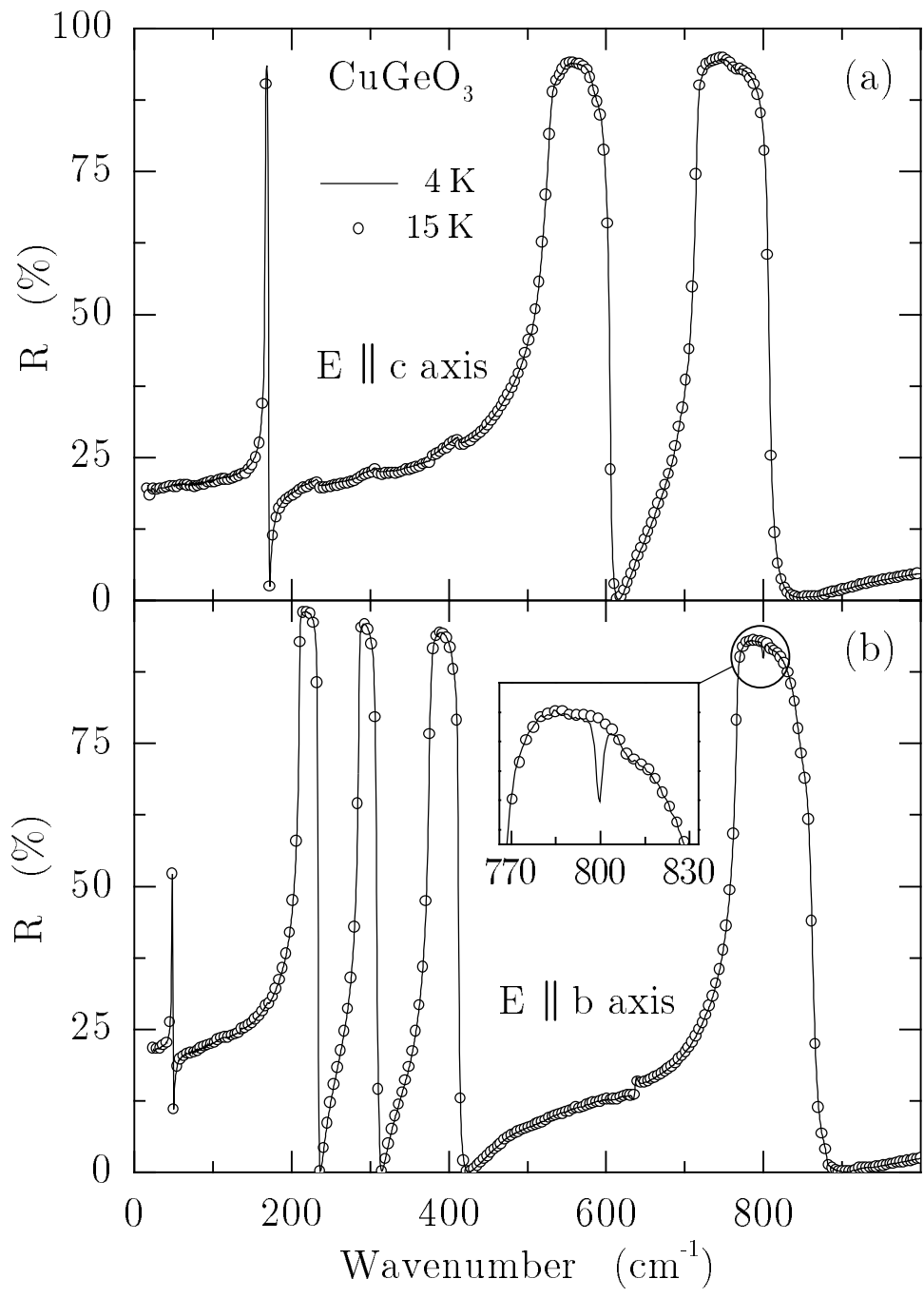
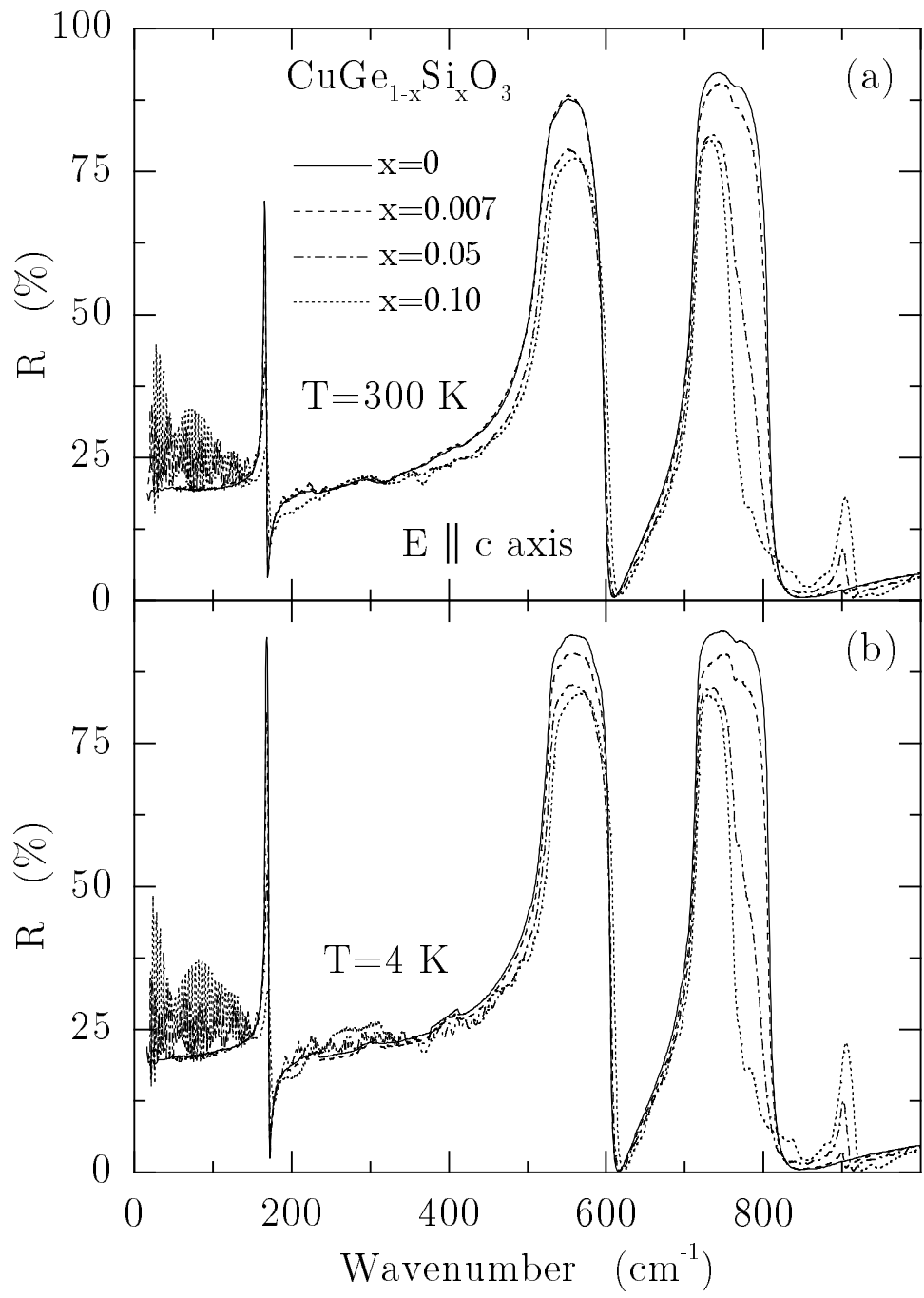
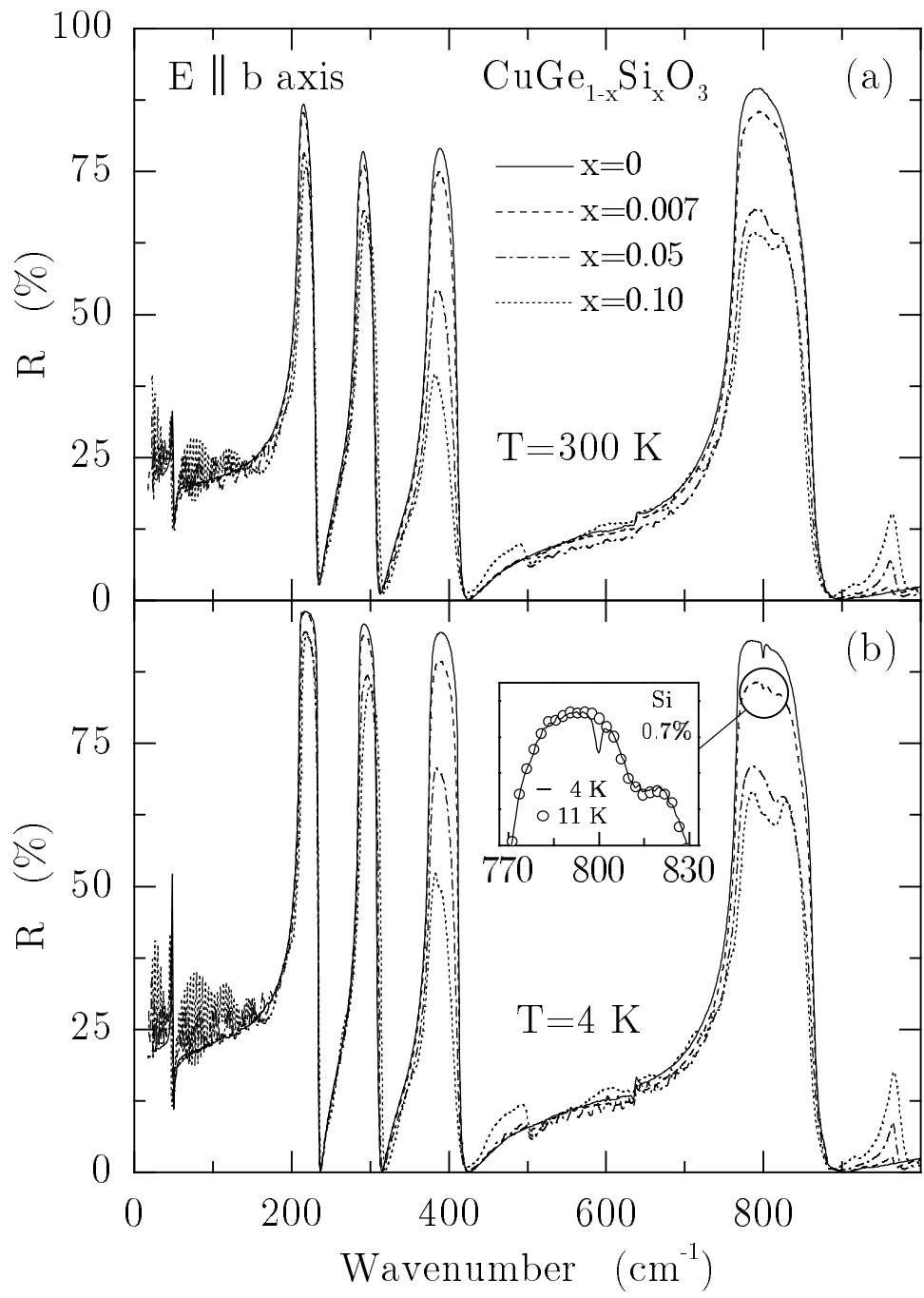
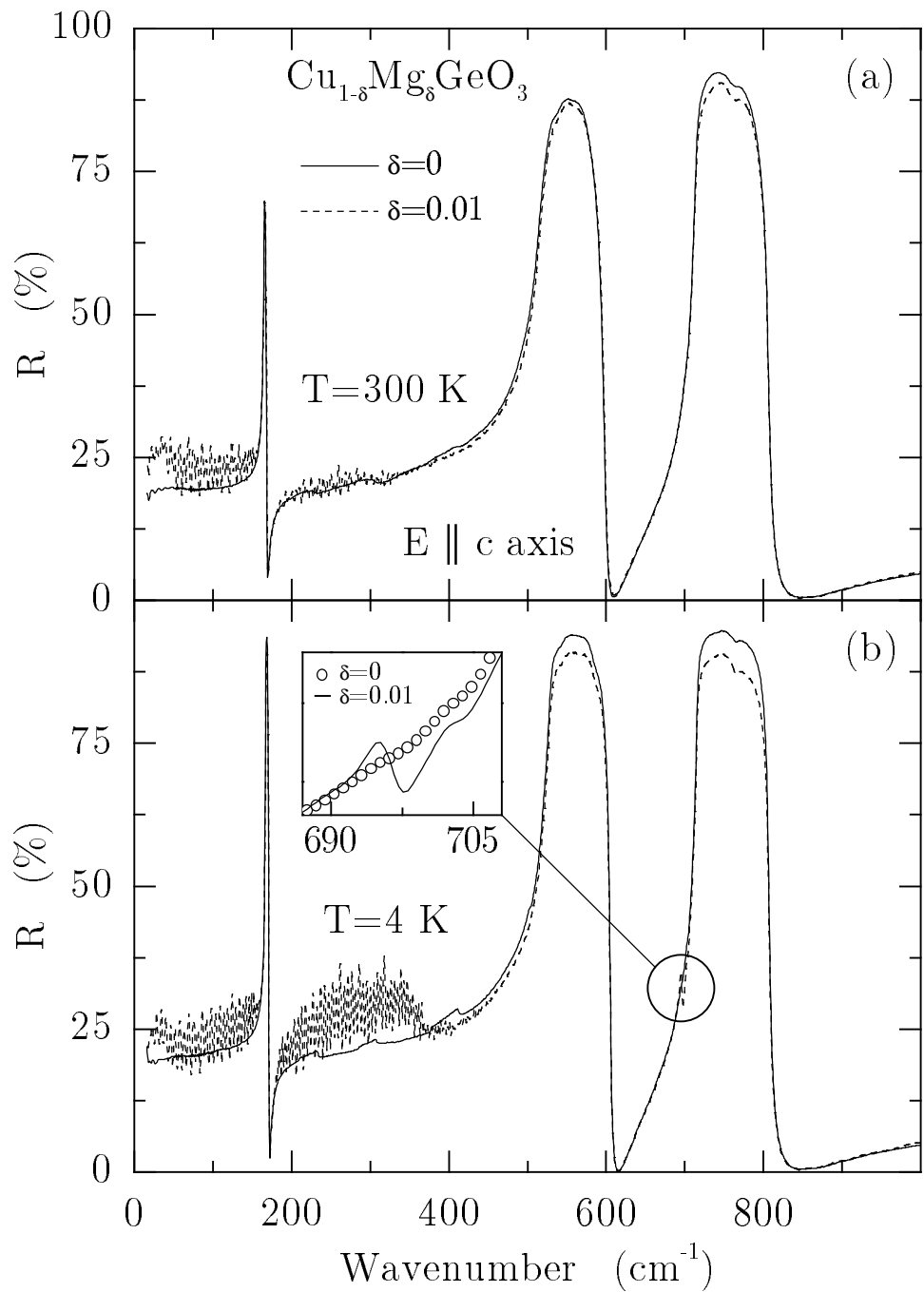
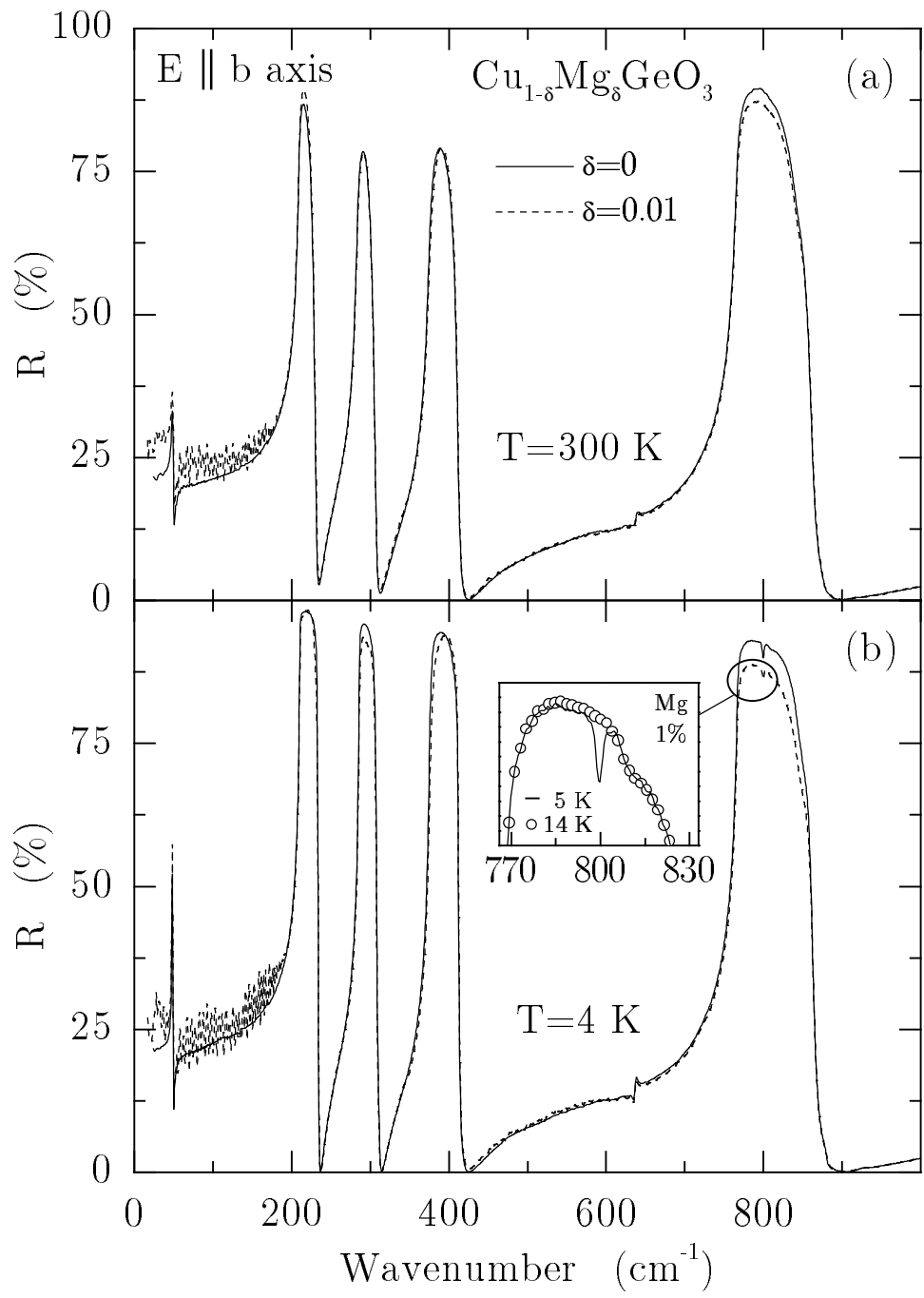
\end{document}